# BAYESIAN DOSE-RESPONSE ANALYSIS FOR EPIDEMIOLOGICAL STUDIES WITH COMPLEX UNCERTAINTY IN DOSE ESTIMATION

Deukwoo Kwon[1,*], F. Owen Hoffman[2], Brian E. Moroz[3], Steven L. Simon[3]

May 29, 2014

[1]Sylvester Comprehensive Cancer Center, University of Miami, Miami, FL

[2]Oak Ridge Center for Risk Analysis, Oak Ridge, TN

[3]Division of Cancer Epidemiology and Genetics, National Cancer Institute, National Institutes of Health, Bethesda, MD

*Address for correspondence: Deukwoo Kwon, Ph.D., Sylvester Comprehensive Cancer Center, University of Miami, 1120 NW 14 Street, Clinical Research Building Rm 1054, Miami, FL 33136.

Abstract

Most conventional risk analysis methods rely on a single best estimate of exposure per person which does not allow for adjustment for exposure-related uncertainty. Here, we propose a Bayesian model averaging method to properly quantify the relationship between radiation dose and disease outcomes by accounting for shared and unshared uncertainty in estimated dose. Our Bayesian risk analysis method utilizes multiple realizations of sets (vectors) of doses generated by a two-dimensional Monte Carlo simulation method that properly separates shared and unshared errors in dose estimation. The exposure model used in this work is taken from a study of the risk of thyroid nodules among a cohort of 2,376 subjects following exposure to fallout resulting from nuclear testing in Kazakhstan. We assessed the performance of our method through an extensive series of simulation tests and comparisons against conventional regression risk analysis methods. We conclude that when estimated doses contain relatively small amounts of uncertainty, the Bayesian method using multiple realizations of possibly true dose vectors gave similar results to the conventional regression-based methods of dose-response analysis. However, when large and complex mixtures of shared and unshared uncertainties are present, the Bayesian method using multiple dose vectors had significantly lower relative bias than conventional regression-based risk analysis methods as well as a markedly increased capability to include the pre-established "true" risk coefficient within the credible interval of the Bayesian-based risk estimate. An evaluation of the dose-response using our method is presented for an epidemiological study of thyroid disease following radiation exposure.

# 1. INTRODUCTION

Proper evaluation of the dose response relationship that takes uncertainty in dose estimation into account is a subject of growing importance in epidemiological research [1-6]. Most epidemiological studies use conventional regression models based on a frequentist paradigm. Those procedures are usually carried out by replacing individual dose estimates with the mean or median dose for all individuals in a cohort who share the same exposure attributes. This assumes that the unknown true values of individual dose will vary at random about unbiased mean doses assigned to individual members of cohort subgroups and that the remaining error structure is completely Berkson [7-11]. Some investigators have used the individual median rather than the individual mean dose in their regression model [12]. Other investigators have estimated the dose-response by treating dose uncertainties as mixtures of Berkson and random classical errors [11, 13-15].

Few studies, however, have addressed the impact of shared (systematic) uncertainties in dose-response analysis. In those cases, methods such as Monte Carlo Maximum Likelihood (MCML) methods [16, 17] have been used. Alternatively, mixtures of shared and unshared/shared random uncertainties have been addressed using Markov chain Monte Carlo (MCMC) methods [18].

Overall, the reported effect of dose uncertainties on the slope of the dose-response by these more sophisticated methods has been rather small, on the order of a factor of two or less [9, 10, 16]. However, the cohorts investigated in those studies, predominantly contained Berkson errors and relatively small amounts of random classical errors and shared uncertainties. There are, however, studies demonstrating that the effect of dose uncertainties on the dose-response can be quite large when the dose uncertainty is large. For example, Kopecky et al. [19] discusses adjustments to the slope of the dose-response for thyroid cancer for a Russian cohort exposed to Chernobyl fallout. The dose uncertainties in individual dose estimates were large (often greater than one order of magnitude about a median estimate) and these uncertainties were assumed to be completely random classical errors. The risk coefficient (ERR $Gy^{-1}$), without accounting for uncertainty in dose estimation, was 48.7 (95% CI=4.8, 1151). When dose uncertainty was taken into account, the risk increased to 138 (95% CI=0.36, $5\times10^8$).

Li et al. [18] found that the slope of the dose response (ERR Gy$^{-1}$) was 11.4 (95% CI =2.1, 59.2) when investigating thyroid neoplasms in the University of Utah cohort exposed to Nevada Test Site fallout [11] using regression calibration based on a single set of individual mean dose estimates and considering only Berkson errors. However, when using Bayesian MCMC and Bayesian Monte Carlo expectation maximization (MCEM) to account for mixtures of unshared and shared Berkson errors and unshared classical errors, Li et al. found that the estimate increased from 11.4 to either 23.1 (95% CI=3.9, 79) or 63.8 (95% CI=5.4, 240), depending on the specific method used.

It is well understood that estimating individual doses in the absence of direct measurements can lead to large and complex uncertainties [20, 21] composed of mixtures of systematic and random errors [22]. The most advanced dose reconstruction methods explicitly separate systematic (i.e., shared) from unshared sources of uncertainty [20, 22-24]. One such method, the two-dimensional Monte Carlo (2DMC), proposed in Simon et al. [22], was used in this work. While any method of generating multiple realizations of possibly true cohort dose distributions could be used with the Bayesian methodology presented here, we are not aware of the existence of dose estimation strategies with this capability other than the 2DMC.

In this work, we use multiple vectors (sets) of cohort doses for two exposure scenarios drawn from a dose reconstruction associated with an epidemiologic study of thyroid disease following radiation exposure [24]: one scenario for external dose with limited shared uncertainty and one scenario for total dose (external + internal) that includes a complex mixture of shared and unshared components. Here, the 2DMC method was used to generate 5,000 vectors of doses for a cohort of 2,376 study subjects. We assessed the performance of five different combinations of a dose-response analysis method and a specific type of dose data set for each of the two exposure scenarios. The two methods assessed were conventional regression and the BMA method which is a Bayesian model averaging (BMA) approach to estimate the dose-response [25]. The performance tests included:

(1) Conventional regression using a single vector of individual mean dose estimates,
(2) Conventional regression using a single vector of individual median dose estimates,
(3) The BMA method using 5,000 vectors of individual dose estimates,
(4) The BMA method using 5,000 vectors of conditional individual mean dose estimates, and
(5) The BMA method using 5,000 vectors of conditional individual median dose estimates.

In the comparison of methods presented here, we use the term "conditional mean" or "conditional median" to indicate a mean or median dose estimate for an individual that has been conditioned on a single sampled value of each dose-model parameter with a value that is shared among members of one or more cohort subgroups, and all other unshared parameter values are resampled 100 times as a basis for obtaining the conditional individual mean or median dose.

Conditional individual mean and median doses are used because the process of deriving the conditioned value reduces or eliminates classical error within each vector of simulated individual doses. To further assess the performance of conditional means and conditional medians, we intentionally introduce additional amounts of classical error for testing with the BMA method.

## 2. METHODS

### 2.1. Dose Estimation Method

The construction of a radiation dosimetry system involves the application of principles of radiation exposure and dose assessment to estimate radiation doses received by individuals from past events [23]. Some studies only require a relatively simple dosimetry system, as is usually the case with controlled exposures in medicine. In those studies, a simple approach using either Monte Carlo or analytical error propagation methods will likely suffice as a means to assess uncertainty since there is often no need to distinguish between shared and unshared errors.

Here, we used a two-dimensional Monte Carlo (2DMC) dosimetry method that is designed to explicitly address random variations and systematic biases in a dosimetry system separately. Upon correct application of the 2DMC method, each dose vector is internally consistent, that is, doses for individuals in a cohort with common exposure attributes are properly correlated with each other. This method is particularly useful for complex dosimetric evaluations of environmental and occupational exposures that are designed to support evaluation of the dose-response relationship [22].

The radiation doses we used for the simulation study are estimates of external and internal radiation dose to the thyroid gland from exposure to radioactive fallout from nuclear testing conducted from 1949-1962 at the former Soviet nuclear weapons test site in Semipalatinsk, Kazakhstan. Exposures due to radioactive fallout were simulated for residents of eight villages in a study of thyroid disease performed by Land et al. [26] but using updated radioactive fallout exposure models [24, 27-29]. The external dose assessment estimates the time-integrated exposure modified by time spent indoors and building shielding characteristics. For external dose, the time of transit of fallout from the detonation site to each of the eight villages as well as the village-and test-specific ground-level exposure rates are important shared errors. The study also estimates the total dose which is a summation of the external dose and the internal dose, the latter which is a quantitative estimate of internal irradiation that arises from consumption of food products contaminated with radioiodines [24, 28]. The internal dose would also have significant shared errors including the ground-level village- and test-specific exposure rate, the parameter describing the interception of fallout by plants (a function of particle size which is related to downwind distance at the site), and the coefficients describing the transfer of radioiodine from feed to each type of animal milk.

For the purpose of testing the performance of the risk analysis methods listed, we simulated 5,000 alternative, possibly true vectors of external and internal doses for the cohort using the 2DMC method. Each dose vector contains a single dose estimate for each subject in the cohort. The 2DMC method was used to ensure that the proper correlations between doses for individuals with common exposure attributes were maintained.

Table 1 of this paper describes the variability and uncertainty in the 5,000 vectors of 2,376 simulated individual doses, conditional individual mean doses and conditional individual median doses produced by the 2DMC algorithm, summarized as a 95% CI. The cumulative distributions of the 2DMC multiple realizations of cohort dose vectors are presented in Fig. 1.

*2.2. Bayesian approach for dose-response with complex dosimetry from a 2DMC algorithm*

Because the 2DMC dose algorithm is able to effectively separate shared from unshared uncertainties in a dose reconstruction, a dose-response methodology is needed to accommodate the multiple realizations of cohort dose vectors. For that reason, we consider an approach for dose-response analysis under the statistical paradigm of model uncertainty, which goes beyond simple measurement error models. In this paper, model uncertainty was assumed to arise from the uncertainty due to dose estimation rather than uncertainty in the shape of the dose-response model. In this work, we have assumed a linear dose-response model.

Here, it is our *a priori* expectation that at least a few of the simulated vectors of doses should be reasonable approximations of the true vector of exposures and, thus, we should be able to evaluate and compare each of the alternative vectors of cohort dose to find the best fit. Due to complexity of the exposure model and the uncertainty of many of the exposure-related parameters, however, we cannot assume that the perfect true vector of doses is included within the multiple realizations of dose produced by the 2DMC dose estimation algorithm. For that reason, we use a Bayesian model averaging (BMA) [25] approach to estimate the dose-response. Some detail about the relationship between our problem and the paradigm of Bayesian model uncertainty is presented in the Appendix.

The BMA method employed in this paper obtains a posterior probability of the slope of the dose-response, by (i) considering all realizations of external dose and total thyroid dose vectors produced by the 2DMC dose reconstruction system, and (ii) evaluating the goodness of fit of each dose vector with the specified disease outcome. Our approach has an advantage over conventional dose-response analysis that use a single dose for each study subject, in that it can address very complex estimations of exposure which include varied patterns of shared and unshared uncertainty in the dose estimation parameters.

Our main goal is to estimate the radiation risk coefficient of a linear dose-response model, $\beta$, and its confidence interval, as well as the credible interval for the Bayesian method, accounting for sampling uncertainty as well as uncertainty in dose estimation.

In this work, we consider a binary disease status variable for thyroid nodules with a logistic regression model and other covariates, the same model formulation as used in Land et al. (2008). For simplicity, we excluded effect modifiers in this paper. We assume, thus, that the probability of disease given covariates $X$ and dose $D$ is given by the expression

$$p = P(Y=1 \mid X, D) = \frac{\exp\left\{\sum_{j=1}^{J} \alpha_j X_j + \log(1+\beta D)\right\}}{1 + \exp\left\{\sum_{j=1}^{J} \alpha_j X_j + \log(1+\beta D)\right\}} \qquad (1)$$

where $X_j$, $j=1,\ldots,J$, are covariates, and $\beta$ is the excess relative risk per Gy (ERR Gy$^{-1}$).

Since uncertainty in dose estimation is represented by K dose vectors, we use a data augmentation approach to deal with multiple realizations of dose vectors. This is implemented in the form of a dose vector selection parameter, $\gamma$, that indicates which dose vector is used in the likelihood for a cohort of N individuals, given by

$$l(\underline{\alpha}, \beta, \gamma \mid y, X, Dose\ vector) = \prod_{i=1}^{N} p_i^{y_i}(1-p_i)^{(1-y_i)}, \qquad (2)$$

In order to perform Bayesian inference, we must formulate prior distributions for all uncertain model parameters. For prior distributions, we use a normal distribution for the $\alpha$ parameters in equation (3) with a large variance in order to be a proper but non-informative prior. For a positive slope of the dose-response, $\beta$, in equation (4), we use an exponential distribution with a large mean. The dose vector selection parameter, $\gamma$, has a multinomial distribution, Multinomial($\underline{\pi}$) in equation (5). The probability vector, $\underline{\pi}$, has a hyper-prior distribution given by a Dirichlet distribution in equation (6), Dirichlet(1,...,1), so that every dose vector has an equal *a priori* probability to be the possibly "true" dose vector.

**Priors:** $\underline{\alpha}=(\alpha_1,...,\alpha_J)$, $\alpha_j \sim N(0, 1000)$, $j=1,2,..J$, (3)

$\beta \sim exponential(100)$, (4)

$\gamma \sim$ Multinomial($\underline{\pi}$), ($\gamma \in \{1,...,K\}$), (5)

$\underline{\pi} = (\pi_1,...,\pi_K) \sim$ Dirichlet(1,...,1) . (6)

The joint posterior distribution is defined as follows:

$p(\underline{\alpha},\beta,\gamma|y,X,Dose) \propto l(\underline{\alpha},\beta,\gamma|y,X,Dose)p(\underline{\alpha})p(\beta)p(\gamma)$ . (7)

The posterior inference is obtained from a Markov chain Monte Carlo (MCMC) calculation. Since the likelihood is formulated with a logistic function, full conditional distributions of the parameters do not have a closed form solution of known distributions. Therefore, the Metropolis-Hastings (MH) method is generally suitable to sample from the posterior distribution of the parameters of interest. Along with the posterior inference for the slope of the dose-response parameter, $\beta$, a relative selection frequency (Bayesian weight) for a specific dose vector is obtained from the posterior distribution of the dose vector index variable, $\gamma$. The Bayesian weight, $p(\gamma = k | y, X, dose\ vector_k) / \sum_{m=1}^{K} p(\gamma = m | y, X, dose\ vector_m)$, $k=1,...,K$, is a relative measure of the goodness-of-fit of each dose vector. For those dose vectors that are relatively close to the unknown vector of true doses, Bayesian weights will typically be relatively larger than those produced for the remaining dose vectors that are less precise approximations of the (unknown) vector of true dose.

In situations when the magnitude of uncertainty in dose estimation is much less than in our example based on the total thyroid dose in the Kazakhstan cohort, we can use either WinBUGS [30] or JAGS [31] for Bayesian computation. For example, Poisson regression and negative binomial regression models were used via JAGS with 1,000 dose vectors that accounted for shared error in Little et al. [32].

When doses are estimated from historical exposure situations and numerous sub-models are necessary to account for the important pathways of exposure, we are likely to have a complex landscape of posterior distributions for the parameters of the dose-response model which, in turn, produces a multimodality of the posterior distribution of the slope of the dose response. In situations such as this that would generally require MCMC methods such as the Metropolis-Hastings (MH) and Gibbs sampling methods, we often face a local-trapping problem which can lead to large biases in parameter estimation and a very narrow confidence interval about the slope, $\beta$. To overcome that problem in our work, we used an advanced MCMC method referred to as a Stochastic Approximation Monte Carlo (SAMC) method [33]. More detail of SAMC is described in the Appendix.

## 3. SIMULATION STUDY

This section describes the process of developing 90 test cases for each of two exposure scenarios in order to assess the performance of the dose-response methods. As noted earlier, the primary difference in the two exposure scenarios is the markedly different amounts of shared and unshared uncertainty in the parameters of the exposure models used to estimate individual dose. The 90 test cases are, simply, 90 different simulated sets of disease status for the entire cohort. Each set is considered to be a "true" set for testing purposes. Each "true" disease set is simulated using a pre-selected vector of "true" dose (one dose per subject) and a pre-specified "true" slope of a linear dose response relationship, along with covariates of the baseline risk from the Kazakhstan epidemiological study [24].

### 3.1. *Specifics of Dose Simulation*

For our Exposure Scenario 1, we used 5,000 vectors of simulated individual thyroid doses from external exposure obtained from the Kazakhstan epidemiologic study [24]. The thyroid dose estimates from external exposure are associated with relatively small amounts of dose uncertainty and small amounts of shared uncertainty contributing to the total uncertainty in individual dose estimation (see Table 1 and Fig 1).

For our Exposure Scenario 2, we used 5,000 vectors of simulated individual <u>total</u> thyroid dose (external + internal exposure). In contrast to Exposure Scenario 1 (external exposures), Exposure Scenario 2 (total exposure) is associated with very large amounts of dose uncertainty, with substantial shared uncertainty contributing to the overall uncertainty in each individual's dose (see Table 1 and Fig. 1).

*3.2.  Simulation of multiple alternative sets of disease status*

For the purposes of performance testing, we simulated multiple alternative realizations of a "true" set of disease status (i.e., the disease status for the 2,376 individuals in the cohort). Each simulated set of disease status was produced using one of the 5,000 vectors of doses for the cohort, one of three pre-specified values of a "true" slope, $\beta$, (ERR $Gy^{-1}$=3,12, and 20), and covariates that affect the baseline risk of thyroid nodules (i.e., age at time of screening and sex). The values for covariates are 2 for age at time of screening, 1.5 for male, and 3 for female. Thus, 5,000 sets of "true" disease status were simulated for each of the three pre-specified "true" value of the slope of the dose response and for each exposure scenario, producing a total of 15,000 simulated sets of "true" disease status per exposure scenario and 30,000 simulated "true" disease status sets overall. Each simulated set of disease status is a potential candidate for performance testing of the two methods of dose response estimation. The following section describes the procedure used to select 180 simulated sets of "true" disease status for use in performance testing to compare the dose response estimates produced by conventional regression methods with the Bayesian method that uses multiple vectors of cohort doses.

*3.3.  Selection of specific disease status sets for performance testing*

While it is feasible to use all 30,000 simulated sets of disease status to test the performance of the conventional regression methods that use a single vector of mean or median doses (Figs. 3 and 4), it is prohibitive to perform 30,000 tests of the Bayesian dose response methods, due to excessive computational requirements. Therefore, we selected a subset of simulated disease status sets to produce a representative range of test conditions that was balanced for subsequent performance testing of the different methods of dose response analysis.

For each of the two exposure scenarios, we selected a subset of 90 sets of disease status, 30 for each of the three pre-specified "true" value of the slope. The selection process we used was as follows:

(i) For each of the three pre-specified "true" values of the slope of the dose response, we simulated 5,000 "true" sets of disease status and estimated the slope of the dose response for each disease status set using conventional regression with individual median doses.

(ii) We ordered the 5,000 estimated slope values (ERR $Gy^{-1}$) in ascending order and divided the ordered sets into ten groups of 500 sets of estimated slope.

(iii) From each of the ten groups of 500 sets of estimated slopes, we selected three at random and identified the corresponding a set of simulated disease status sets to produce 30 "true" simulated sets of disease status for each of three pre-specified "true" values of the slope, for a total of 90 sets of disease status selected per test scenario (180 sets of disease status selected overall among both test scenarios).

In the selection procedure described above, the subset of 30 disease status sets are representative of the overall range of disease status sets and are symmetrically balanced about the specific disease status set that has an estimated slope equal to the "true" slope. This balance avoids a bias in the selection of sets of disease status which have estimated slopes (using conventional regression with individual median doses) towards over-or under-estimation of the "true" slope.

The selection procedure essentially involves pre-testing conventional regression using median doses with all 30,000 simulated sets of disease status. Thus, our reported test results for this method are effectively pre-determined and will have a pre-determined high rate of successful outcomes. About 50% of the disease vectors selected for testing in Exposure Scenario 2 will have corresponding 95% confidence intervals that included the prescribed "true" slope (See Table 2). Therefore, the test results for conventional regression using median doses should be considered a baseline when comparing the test results obtained with other methods of dose response evaluation.

*3.4.  Comparison of performance of Bayesian procedure and conventional regression models*

In the performance testing conducted in this paper, two basic types of statistical methods were used to determine the dose-response of the simulated "true" sets of disease status: (1) conventional regression using a single vector of individual doses, and (2) Bayesian model averaging (BMA) using multiple dose vectors produced from the 2DMC dose estimation algorithm. In our analysis with the conventional regression-based method, we used a single vector of individual mean or individual median doses obtained from 5,000 dose realizations for each member of the cohort.

For each simulation of a disease status set, one cohort dose vector out of the 5,000 simulated dose vectors produced by the 2DMC dose estimation algorithm was pre-selected as a "true" cohort dose vector. This "true" dose vector used to simulate a selected set of disease status is excluded from the full 5,000 dose vectors prior to use of the BMA method for dose response analysis, reflecting the condition that the true exposure model is not included in our model uncertainty paradigm (see Appendix). The BMA method thus uses 4,999 dose vectors to evaluate the dose response for a given "true" set of simulated disease status.

Performance testing consisted of comparing the results of dose-response estimation produced by conventional regression and the BMA method for each of the simulated "true" sets of disease status selected for each of two test scenarios. These tests were conducted to determine whether the value of the "true" slope was captured within (i) the 95% confidence interval (CI) of the slope parameter, $\beta$, estimated from the conventional regression model, and (ii) the Bayesian credible interval of the posterior distribution of the slope parameter, $\beta$, estimated from the BMA method. Bayesian parameter estimates and their corresponding 95% Bayesian credible interval were obtained based on 40,000 posterior samples after 10,000 burn-in iterations (*burn-in* is an initial set that is not used for inferences). Hereafter, we used the notation CI for both confidence interval and Bayesian credible interval.

*3.5.  Multiple vectors of conditional mean and median dose*

Because of our concern that each of the simulated cohort dose vectors from the 2DMC dose estimation procedure might inadvertently over-estimate inter-individual variability of true doses and, thus, introduce non-intended classical error into the analysis, multiple conditional individual mean doses and conditional individual median doses were also produced. A vector of conditional individual mean and median doses was obtained (separately) after resampling 100× all parameters used for dose-estimation specified as representing unshared uncertainty in each subject's dose. This resampling was "conditioned" on a fixed set of dose-model parameter values that were shared among members of cohort subgroups. Selection of a fixed set of shared parameter values and resampling 100× all unshared parameters for each of all 2,376 individuals in the cohort was repeated 5,000 times to produce 5,000 vectors of conditional individual mean and median doses. Each vector of conditional individual mean or median doses is assumed to be a possibly "unbiased" central estimate of doses for subgroups of individuals sharing the same exposure attributes; all remaining uncertainty is attributed to random variation of the unknown individual true dose about the possibly unbiased central value of dose assigned to members of the subgroup (i.e., all remaining uncertainty is assumed to be 100% Berkson).

## 4. RESULTS

### 4.1. Inclusion percentage of the "true" test slope within the 95% CI slope estimates

As one indicator of performance of each of the methods of dose response, we used the percentage with which the pre-specified "true" slope was captured by the 95% CI of the computed slope estimate (Table 2). Since conventional regression with the single vector of individual median doses was used to select the set of simulated disease status used for testing, the test results obtained with this method should be considered as a benchmark against which the other methods are compared.

### 4.2. Testing of conventional regression analysis methods with a single vector of mean doses against those using median doses

As part of performance testing, we first compared conventional regression using a single vector of individual mean doses with the same method using a vector of median doses. For both exposure scenarios, the inclusion percentages for the single vector of individual mean doses were lower than those for the individual median doses though the difference was small, 91.1% for the mean doses compared with 94.4% for the median doses (Table 2, Fig. 4, and Supplementary Fig. 1). However, for Exposure Scenario 2, the difference in the inclusion percentage between these two data sets was dramatic, 12.2% for means compared to 51.1% for medians. There was also strong evidence of a marked systematic bias in the dose response estimated by the individual mean doses towards underestimation of the "true" slope of the test data (Fig. 5, also see Supplementary Fig. 2).

In order to examine the representativeness of the selected disease status sets used for testing, we calculated inclusion percentages of the "true" value of the slope with conventional regression using single vectors of mean and median doses for all generated disease status sets. Inclusion percentages are shown in parentheses in Table 2. Inclusion percentages obtained for the subset of 30 selected disease sets were comparable to those obtained for all 5,000 disease sets simulated using a "true" value of the slope.

We investigated possible reasons why conventional regression with the mean doses showed an apparent inferior performance compared with the results obtained with the median doses in Exposure Scenario 2 (Fig. 3, also see Supplementary Fig. 2). In Figs. 2 and 3, the slopes (ERR $Gy^{-1}$) for all 5,000 disease sets were estimated using both mean doses and median doses. While the slope estimates with median doses were symmetrically distributed among the multiple disease status sets for both Exposure Scenario 1 and 2, slope estimates using mean doses were highly skewed in Exposure Scenario 2. The skewness for the mean doses seen for Exposure Scenario 2 (Figs. 3, 5, and Supplementary Fig. 2) is the result of very high uncertainty in each subject's dose, with the 5,000 dose estimates for each study subject approximating a lognormal distribution with about 30% of the cohort having GSDs greater than 3.0 and with 5% having GSDs greater than 4.0. This high uncertainty in each individual's total thyroid dose is the consequence of both shared and unshared sources of uncertainty contributing to the distribution of dose estimates for each person.

With a wide distribution of dose estimates for each individual that approximates a lognormal distribution, the mean of that distribution will be a much larger value than the median. Because this wide distribution is the outcome of a large combination of multiple sources of dosimetric uncertainty (in addition to random variability of true dose among individual members of cohort subgroups sharing the same exposure attributes), the individual mean dose for members of any subgroup will be biased high with respect to the true mean dose for that subgroup. Thus, when a single vector of individual mean doses is obtained from 5,000 dose realizations per study subject, and when these dose realizations contain a combination of major sources of uncertainty in addition to random inter-individual variability of true doses, conventional regression using this vector of mean doses will produce a systematic under-estimation of the majority of the "true" test slopes of the dose response.

*4.3.    Comparison of conventional regression using median doses with Bayesian methods*

When we compared the performance results obtained with the single vector of cohort median doses against the three Bayesian model averaging approaches for Exposure Scenario 1 (Table 2), we found equally high inclusion percentages (95.6%) for all three Bayesian methods, comparable to the conventional regression with median doses (94.4%). These high inclusion percentages reflect the relatively small amounts of uncertainty associated with the external thyroid dose estimates in the Kazakhstan study [24].

In contrast, for Exposure Scenario 2 (total dose), the three Bayesian methods showed substantially higher inclusion percentages (70 to 91%) than did conventional regression using the single vector of cohort median doses (52.2%). The Bayesian approach using multiple cohort vectors of conditional individual mean doses and multiple vectors of conditional individual median doses produced very high inclusion percentages of 88.9 and 91.1%, respectively. These results were substantially improved over the Bayesian approach using simulated individual doses (70%).

In about 9% of the cases where the Bayesian approach using the conditional individual median dose vectors did not contain the "true" test slope of the dose response within its 95% CI, the other dose response methods failed as well. We believe that these cases may represent special situations with low statistical power to detect an underlying true effect.

*4.4.    Evaluation of the relative bias of the slope estimates*

We also computed the relative bias associated with the estimate of the slope of the dose response in each test (Table 3). Relative bias is calculated as absolute bias (estimated slope – "true" slope) divided by the "true" slope. The performance of the various dose-response evaluation methods can be compared using the magnitude of relative biases. In Exposure Scenario 1, the magnitudes of relative biases were small for all methods of dose response, a finding that is consistent with the high inclusion percentages in Table 2.

However, in Exposure Scenario 2, the Bayesian method with conditional individual median dose vectors showed the smallest relative bias (0.29). This can be compared with 0.31 obtained with the Bayesian approach using conditional mean doses and with 0.40 obtained with the Bayesian approach using simulated individual doses. In contrast, the conventional regression analyses using a single vector of mean doses and median doses produced an overall relative bias of 0.66 and 0.46, respectively. In Exposure Scenario 2, most of the relative bias for conventional regression using mean doses is due to a systematic tendency to overestimate the "true" test slope, while conventional regression using median doses results, as expected, in a symmetrical over- and under-estimate of the "true" test slope (Figs. 4 and 5 and Supplementary Figs. 1 and 2). This expectation arises because conventional regression using individual median doses was used to select the subset of simulated disease vectors used for testing.

### *4.5. Evaluation of the width of the 95% CI's*

Here we discuss the magnitude of uncertainty associated with the various estimates of the slope of the dose response using the half-width of 95% CI (Table 3). Exposure Scenario 1 contains relatively small differences between the width of the CI between conventional regression and any of the Bayesian methods. However, in Exposure Scenario 2, the difference between the half-width of the CI for conventional regression using a single vector of median doses and the Bayesian approach using multiple dose vectors was more than double (3.8 compared with 8.1). In addition, the half-width of the CI comparing conventional regression using mean doses to conventional regression using median doses was also more than double (1.72 compared with 3.8). The larger values for the half width of the CI for the Bayesian methods reflects wider CI's that contain the influence of uncertainty in dose estimation as well as uncertainty in the fit of the dose response to the disease status set.

Interestingly, the half-width of the CI for conventional regression with a single vector of mean doses decreased in Exposure Scenario 2 compared with Exposure Scenario 1. In Exposure Scenario 2, because of the presence of very large uncertainties in the simulated individual doses, the values of the individual mean doses are larger than the values of individual median doses. These larger individual mean dose estimates most often produced under- estimates of the "true" test slope of the dose-response (Table 1). This caused the half-width of the 95% CI of the estimate of the slope to shrink in size.

We also examined the relative width of the ratio of the upper CI of the slope estimates to the estimate of the slope (Table 2). These relative widths were very stable across different values of the test slope. The relative width of the upper CI from the Bayesian methods were about 6% and 35% larger in Exposure Scenarios 1 and 2, respectively, than those produced by the conventional regression analysis with median or mean doses. As with the results for the half-width of the CI, larger values of the relative width of the upper bound of the CI for the Bayesian analyses, reflects uncertainty in dose estimation as well as uncertainty in the fit of the dose response model to the test disease status set. The effect of dose uncertainty is not included in the 95% CI's produced by conventional regression analysis using a single vector of individual mean or individual median doses.

*4.6.    Evaluation of simulated classical errors*

For the conventional regression analysis methods, the presence of classical error in dose estimation has been demonstrated to bias the estimate of the slope of the dose response towards the null [4]. In this paper, we examined the effect of classical error on the dose response determined by the Bayesian method. Additional amounts of classical error were intentionally introduced, via simulation, into the 5,000 2DMC dose vectors using a multiplicative lognormal classical error factor. This factor was defined by a GM of 1.0 with increasing values of the geometric standard deviation (GSD) to simulate increasing amounts of unshared random variability in dose estimation due to random factors other than true inter-individual variability of true dose.

The lognormal classical error factor was applied within the unshared part of the 2DMC dose calculation for the total thyroid dose. For the Bayesian method using 4,999 vectors of simulated individual doses, the lognormal factor was sampled once for each of individual doses within a dose vector. This was repeated per individual for each successive dose vector. For the Bayesian approach with 4,999 vectors of conditional individual mean and median doses, the lognormal classical error factor was applied per individual within each of the conditional 100× resampling of unshared parameters. This procedure inflated the variability of each individual dose distribution within each conditional dose vector.

Values of the GSD for the lognormal classical error simulation factor ranged from 1.0 (no classical error), to 1.3 (low), 1.5 (modest), 2 (high), and 3 (very high). In Fig. 6 we show how the slope estimates and corresponding 95% CIs are affected by increasing amounts of classical error. All Bayesian approaches to dose response analysis showed increasing attenuation in the estimated slope of the dose response as the amount of simulated classical error increased, except the Bayesian approach using multiple vectors of conditional individual median doses. The response to the simulation of multiplicative classical error was most pronounced with the Bayesian approach using multiple vectors of simulated individual doses.

In summary, under conditions of high and complex dose uncertainty (as represented by the total thyroid doses in Exposure Scenario 2), all three Bayesian approaches performed markedly better in our simulation tests than did the more conventional regression analysis using a single vector of mean or median doses. However, all five approaches performed very well when uncertainty in individual doses was low and when shared sources of uncertainty were small. Application of these methods to an actual set of individuals with thyroid nodules and thyroid cancer is demonstrated in Land et al. [24].

## 5. Application to the Actual Prevalence of Thyroid Nodules

The above tests were based on simulation of the prevalence of thyroid nodules that reflected a pre-determined "true" slope of the dose response. To demonstrate the utility of the Bayesian model averaging method with multiple realizations of cohort doses, we proceeded to apply this method to the actual observed cases of thyroid nodules in this cohort for which there were 177 cases in males and 571 cases in females among 2,376 individuals [24].

As in Land et al. [26], we used same model as follows:

$$\text{Odds} = \exp[\sum \alpha_i X_i] \times (1 + \sum \beta_j Y_j \times \exp\{\sum \gamma_k Z_k\})$$

The subscripted Greek letters $\alpha_i$, $\beta_j$ and $\gamma_k$ denote unknown parameters, and the corresponding subscripted capital letters $X_i$, $Y_j$ and $Z_k$ denote potential risk factors, radiation dose (external and internal doses given separately), and gender variable (-1 for male and 1 for female) for effect modifiers, respectively. The exponential expression $\exp\{\alpha_i X_i\}$ represents the baseline odds, i.e., when the radiation dose is zero. The odds ratio (OR) is the ratio of the odds to the baseline odds,

$$OR = 1 + \sum \beta_j Y_j \times \exp\{\sum \gamma_k Z_k\}$$

and the excess odds ratio (EOR) is the odds ratio minus 1.

The previous 2008 study [26], which used conventional regression with a single vector of "best estimate" individual thyroid doses, showed that the slope of the dose response (EOR $Gy^{-1}$) for external exposure was higher (prior to modifying by sex), but not statistically significantly different from internal exposure. We get similar results to those reported in 2008 study, using conventional dose-response analysis with a single vector of arithmetic mean dose per person. However, when a single vector of individual median doses is used with conventional regression, the central estimate of the slope of the dose response for internal exposures is about a factor of 5 higher than that obtained using the single vector of "best estimate" doses from the 2008 paper [26].

In contrast to conventional regression-based methods with a single vector of individual dose estimates, our Bayesian model averaging method produced risk estimates for internal exposures that were equal to or higher than the risk estimates for external exposures. However, because of the large overlap of 95% CIs, differences in risk estimates between internal and external exposure were not statistically significant. Unlike the 2008 results [26], the dose responses for external exposure (prior to modification by sex) were not statistically significant (24). Central estimates of the EOR $Gy^{-1}$ for external exposure prior to modification by sex (1.5 to 1.6) were comparable among all Bayesian analyses.

The Bayesian model averaging method did confirm the presence of a significantly larger EOR $Gy^{-1}$ for males than for females. Previously in 2008 [26] this difference was reported to be a factor of about 11, with the EOR $Gy^{-1}$ for males being larger than that for females. In the present analysis, the EOR $Gy^{-1}$ for males, 9.99 (95% CI 2.33, 19.1) increased by almost a factor of 30 over that for females, 0.35 (95% CI, 0.000011, 1.0). The above results are based on Bayesian model averaging with multiple (5,000) vectors of conditional individual median doses since we this approach performed the best in our simulation tests compared with Bayesian model averaging using multiple vectors of simulated individual doses or with conditional individual mean doses. Additional description of this cohort and further details of the application conventional regression analysis and Bayesian model averaging with multiple cohort dose vectors to evaluate the dose response of the actual prevalence of thyroid nodules can be found in Land et al. [24].

## 6. CONCLUSIONS

We have demonstrated the use of a Bayesian-based risk analysis method that can be used with multiple realizations of possibly true cohort doses produced from a dosimetry system, e.g., the 2DMC dose reconstruction method, which explicitly accounts for shared and unshared errors in dose reconstruction. We have shown that unlike simpler and more traditional frequentist-based dose-response calculation strategies, e.g., regression calibration, that are adequate when shared errors are small, this Bayesian method gives more reliable and robust results when the shared error among study subjects is substantial. The Bayesian analysis with conditional individual median doses captured the "true" slope of the dose response without attenuation, even when there were large amounts of simulated classical error introduced into the dose estimates. For this reason, we recommend that the Bayesian approach using multiple vectors of conditional individual median doses as our preferred method for quantifying the dose response when dose uncertainties are large and when the amount of uncertainty shared across cohort subgroups is substantial.

Application of our Bayesian method to the actual prevalence of thyroid nodules in the Kazakhstan nuclear test site cohort demonstrated marked differences from what was reported previously using conventional regression with a single vector of dose estimates. The risk per unit dose from internal exposure increased by about a factor of six and the dependency of risk on sex increased from a previously reported factor of 11 to nearly 30 with males having higher excess odds ratios per unit dose than females. We believe these results to be more reliable because of the superior capabilities of the Bayesian Averaging Method as demonstrated through the extensive simulation tests performed and discussed in Section 4. This represents the first application of multiple realizations of uncertain cohort doses combined with a Bayesian Model Averaging method to evaluate risk in an exposed cohort exposed to ionizing radiation.

Clearly there are special cases when dose uncertainty is relatively small and where sources of shared uncertainty are minor, as is the case for external doses for the Kazakhstan study cohort. Under those situations, the conventional dose-response method using a single vector of individual median doses produced results comparable to the Bayesian approaches. Even when dose uncertainties were small, conventional dose-response estimation using a single vector of individual median doses performed marginally better than with a single vector of individual mean doses. Thus, the primary utility of conventional dose-response estimation using a single vector of individual median doses appears to be in situations where (a) dose uncertainty is relatively small and (b) dose uncertainty is mainly the result of random unshared errors among individuals, with only a minor contribution of shared sources of dose uncertainty.

Radiation dosimetry is but one type of exposure analysis where substantial individual and shared errors may be involved. Exposures due to other occupational hazards and, occasionally, due to exposure that affect members of the public, may also be highly uncertain. The methods, principles, and findings discussed here should, in principle, equally apply well, whenever uncertainty in an epidemiologic study is an important concern.

# REFERENCES

1. DeKlerk NH, English DR, Armstrong BK. A review of the effects of random measurement error on relative risk estimates in epidemiological studies. *Int J Epidem* 1989; **18**: 705-712.

2. Armstrong BG. Effect of measurement error on epidemiological studies of environmental and occupational exposures. *Occup Environ Med* 1998; **55**: 651-656.

3. Ron E, Hoffman FO. Uncertainties in radiation dosimetry and their impact on dose-response analysis. In Uncertainties in radiation dosimetry and their impact on dose-response analysis, Editor (ed)^(eds). NIH: City, 1997.

4. Carroll RJ, Ruppert D, Stefanski LA, Crainiceanu CM. *Measurement Error in Nonlinear Models: A Modern Perspective.* (Second edn). Chapman & Hall/CRC: Boca Raton, FL, 2006.

5. Schafer DW, Gilbert ES. Some statistical implications of dose uncertainty in radiation dose-response analyses. *Radiat Res* 2006; **166**: 303-312.

6. National Council on Radiation Protection and Measurements (NCRP). Uncertainties in the Measurement of Radiation Risks and Probability of Causation. In Uncertainties in the Measurement of Radiation Risks and Probability of Causation, Editor (ed)^(eds). City, 2012.

7. Lagarde F, Pershagen G, Akerblom G, Axelson O, Baverstam U, Damber L, Enflo A, Svartengren M, Swedjemark GA. Residential radon and lung cancer in Sweden: risk analysis accounting for random error in the exposure assessment. *Health Phys* 1997; **72**: 269-276.

8. Stram DO, Langholz B, Huberman M, Thomas DC. Correcting for exposure measurement error in a reanalysis of lung cancer mortality for the Colorado Plateau Uranium Miners cohort. *Health Phys* 1999; **77**: 265-275.

9.	Lubin JH, Schafer DW, Ron E, Stovall M, Carroll RJ. A reanalysis of thyroid neoplasms in the Israeli tinea capitis study accounting for dose uncertainties. *Radiat Res* 2004; **161**: 359-368.

10.	Darby S, Hill D, Deo H, Auvinen A, Barros-Dios JM, Baysson H, Bochicchio F, Falk R, Farchi S, Figueiras A, Hakama M, Heid I, Hunter N, Kreienbrock L, Kreuzer M, Lagarde F, Makelainen I, Muirhead C, Oberaigner W, Pershagen G, Ruosteenoja E, Rosario AS, Tirmarche M, Tomasek L, Whitley E, Wichmann HE, Doll R. Residential radon and lung cancer--detailed results of a collaborative analysis of individual data on 7148 persons with lung cancer and 14,208 persons without lung cancer from 13 epidemiologic studies in Europe. *Scand J Work Environ Health* 2006; **32 Suppl 1**: 1-83.

11.	Lyon JL, Alder SC, Stone MB, Scholl A, Reading JC, Holubkov R, Sheng X, White GL, Jr., Hegmann KT, Anspaugh L, Hoffman FO, Simon SL, Thomas B, Carroll R, Meikle AW. Thyroid disease associated with exposure to the Nevada nuclear weapons test site radiation: a reevaluation based on corrected dosimetry and examination data. *Epidemiology* 2006; **17**: 604-614.

12.	Kopecky KJ, Davis S, Hamilton TE, Saporito MS, Onstad LE. Estimation of thyroid radiation doses for the hanford thyroid disease study: results and implications for statistical power of the epidemiological analyses. *Health Phys* 2004; **87**: 15-32.

13.	Mallick B, Hoffman FO, Carrol RJ. Semiparametric regression modeling with mixtures of Berkson and classical error, with application to fallout from the Nevada test site. *Biometrics* 2002; **58**: 13-20.

14.	Lubin JH, Wang ZY, Wang LD, Boice JD, Jr., Cui HX, Zhang SR, Conrath S, Xia Y, Shang B, Cao JS, Kleinerman RA. Adjusting lung cancer risks for temporal and spatial variations in radon concentration in dwellings in Gansu Province, China. *Radiat Res* 2005; **163**: 571-579.

15. Pierce DA, Vaeth M, Cologne JB. Allowance for random dose estimation errors in atomic bomb survivor studies: a revision. *Radiat Res* 2008; **170**: 118-126.

16. Stayner L, Vrijheid M, Cardis E, Stram DO, Deltour I, Gilbert SJ, Howe G. A Monte Carlo maximum likelihood method for estimating uncertainty arising from shared errors in exposures in epidemiological studies of nuclear workers. *Radiat Res* 2007; **168**: 757-763.

17. Kesminiene A, Evrard AS, Ivanov VK, Malakhova IV, Kurtinaitise J, Stengrevics A, Tekkel M, Chekin S, Drozdovitch V, Gavrilin Y, Golovanov I, Kryuchkov VP, Maceika E, Mirkhaidarov AK, Polyakov S, Tenet V, Tukov AR, Byrnes G, Cardis E. Risk of thyroid cancer among chernobyl liquidators. *Radiat Res* 2012; **178**: 425-436.

18. Li Y, Guolo A, Hoffman FO, Carroll RJ. Shared uncertainty in measurement error problems, with application to Nevada Test Site fallout data. *Biometrics* 2007; **63**: 1226-1236.

19. Kopecky KJ, Stepanenko V, Rivkind N, Voilleque P, Onstad L, Shakhtarin V, Parshkov E, Kulikov S, Lushnikov E, Abrosimov A, Troshin V, Romanova G, Doroschenko V, Proshin A, Tsyb A, Davis S. Childhood thyroid cancer, radiation dose from Chernobyl, and dose uncertainties in Bryansk Oblast, Russia: a population-based case-control study. *Radiat Res* 2006; **166**: 367-374.

20. Hoffman FO. *Environmental Dose Reconstruction: How large can uncertainty be when models take the place of measurements? In: Uncertainties in Radiation Dosimetry and Their Impact on Dose-Response Analysis* National Cancer Institute/National Institutes of Health Publication, 1999.

21. Hoffman FO, Ruttenber AJ, Apostoaei AI, Carroll RJ, Greenland S. The Hanford Thyroid Disease Study: an alternative view of the findings. *Health Phys* 2007; **92**: 99-111.

22. Simon SL, Hoffman FO, Hofer E. The Two-Dimensional Monte Carlo: A New Methodologic Paradigm for Dose Reconstruction for Epidemiological Studies. *Submitted to Radiation Research* 2014.

23. National Council on Radiation Protection and Measurements (NCRP). Radiation Dose Reconstruction: Principles and Practices. In Radiation Dose Reconstruction: Principles and Practices, Editor (ed)^(eds). City, 2009.

24. Land C, Kwon D, Hoffman F, Moroz B, Drozdovitch V, Bouville A, Beck H, Luckyanov N, Weinstock RM, SL S. Accounting for shared and unshared dosimetric uncertainties in the dose-response for ultrasound-detected thyroid nodules following exposure to radioactive fallout. 2014.

25. Hoeting JA, Madigan D, Raftery AE, Volinsky CT. Bayesian Model Averaging: A Tutorial. *Statistical Science* 1999; **14**: 382-417

26. Land CE, Zhumadilov Z, Gusev BI, Hartshorne MH, Wiest PW, Woodward PW, Crooks LA, Luckyanov NK, Fillmore CM, Carr Z, Abisheva G, Beck HL, Bouville A, Langer J, Weinstock R, Gordeev KI, Shinkarev S, Simon SL. Ultrasound-detected thyroid nodule prevalence and radiation dose from fallout. *Radiat Res* 2008; **169**: 373-383.

27. Gordeev K, Shinkarev S, Ilyin L, Bouville A, Hoshi M, Luckyanov N, Simon SL. Retrospective dose assessment for the population living in areas of local fallout from the Semipalatinsk Nuclear Test Site Part II: Internal exposure to thyroid. *J Radiat Res* 2006; **47 Suppl A**: A137-141.

28. Gordeev K, Shinkarev S, Ilyin L, Bouville A, Hoshi M, Luckyanov N, Simon SL. Retrospective dose assessment for the population living in areas of local fallout from the Semipalatinsk nuclear test site Part I: External exposure. *J Radiat Res* 2006; **47 Suppl A**: A129-136.

29. Simon SL, Beck HL, Gordeev K, Bouville A, Anspaugh LR, Land CE, Luckyanov N, Shinkarev S. External dose estimates for Dolon village: application of the U.S./Russian joint methodology. *J Radiat Res* 2006; **47 Suppl A**: A143-147.

30. Spiegelhalter D, Thomas A, Best N. *WinBUGS Version 1.4 User Manual.* Medical Research Council Biostatistics Unit: Cambridge, UK, 2003.

31. Plummer M. JAGS : A program for analysis of Bayesian graphical models using Gibbs sampling. In JAGS : A program for analysis of Bayesian graphical models using Gibbs sampling, Editor (ed)^(eds). City, 2003.

32. Little MP, Kwon D, Doi K, Simon SL, Preston DL, Doody MM, Lee T, Miller JS, Kampa DM, Bhatti P, Tucker JD, Linet MS, Sigurdson AJ. Association of chromosome translocation rate with low dose occupational radiation exposures in US radiologic technologists. *Radiat Res* 2014 (In press).

33. Liang F, Liu C, Carroll RJ. Stochastic Approximation in Monte Carlo Computation. *Journal of the American Statistical Association* 2007; **102**: 305-320.

Table 1. Estimates of the distribution of individual doses in the Kazakhstan cohort [from Land et al. (2014)]

**Exposure Scenario 1: External thyroid dose**

| Cohort dose | Individual mean values[a] | Individual median values[b] | 2DMC Individual Dose Realizations[c] | | 2DMC Conditional Individual Mean Doses[d] | | 2DMC Conditional Individual Median Doses[e] | |
|---|---|---|---|---|---|---|---|---|
| | | | 95% CI | | 95% CI | | 95% CI | |
| Minimum(Gy) | 0.0E+00 | 0.0E+00 | 0.0E+00 | 0.0E+00 | 0.0E+00 | 0.0E+00 | 0.0E+00 | 0.0E+00 |
| Maximum (Gy) | 5.4E-01 | 5.0E-01 | 4.4E-01 | 1.3E+00 | 3.3E-01 | 8.6E-01 | 3.1E-01 | 8.2E-01 |
| Median (Gy) | 3.7E-02 | 3.5E-02 | 2.1E-02 | 3.7E-02 | 2.3E-02 | 4.6E-02 | 2.2E-02 | 4.3E-02 |
| Mean (Gy) | 5.6E-02 | 5.2E-02 | 4.4E-02 | 7.1E-02 | 4.4E-02 | 7.1E-02 | 4.1E-02 | 6.7E-02 |
| Variance (Gy$^2$) | 4.5E-03 | 3.7E-03 | 3.2E-03 | 1.0E-02 | 2.6E-03 | 8.3E-03 | 2.3E-03 | 7.3E-03 |

**Exposure Scenario 2: Total thyroid dose**

| Cohort dose | Individual mean values[a] | Individual median values[b] | 2DMC Individual Dose Realizations[c] | | 2DMC Conditional Individual Mean Doses[d] | | 2DMC Conditional Individual Median Doses[e] | |
|---|---|---|---|---|---|---|---|---|
| | | | 95% CI | | 95% CI | | 95% CI | |
| Minimum(Gy) | 0.0E+00 | 0.0E+00 | 0.0E+00 | 0.0E+00 | 0.0E+00 | 0.0E+00 | 0.0E+00 | 0.0E+00 |
| Maximum (Gy) | 4.8E+01 | 1.2E+01 | 1.9E+00 | 1.5E+02 | 8.8E-01 | 4.5E+01 | 7.1E-01 | 3.6E+01 |
| Median (Gy) | 1.6E-01 | 1.0E-01 | 7.0E-02 | 1.6E-01 | 8.3E-02 | 2.4E-01 | 6.6E-02 | 1.6E-01 |
| Mean (Gy) | 4.5E-01 | 1.7E-01 | 1.2E-01 | 2.2E+00 | 1.2E-01 | 2.3E+00 | 9.5E-02 | 1.2E+00 |
| Variance (Gy$^2$) | 5.6E-01 | 3.7E-02 | 2.5E-02 | 8.1E+01 | 1.6E-02 | 3.3E+01 | 9.7E-03 | 1.0E+01 |

A comparison of cohort dose estimates using a single vector of 2376
 [a] individual mean doses and
 [b] individual median doses, with a 2DMC dose estimation producing 5000 vectors of 2376
 [c] simulated individual doses
 [d] conditional individual mean doses, and
 [e] conditional individual median doses.
The single vector of individual mean and median doses are obtained from 5000 dose realizations per each of 2376 study subjects.

Table 2. Inclusion percentage (%) of "true" value of the slope (ERR/Gy) summarized from 90 simulated "true" sets of disease status per exposure scenario.

| | Exposure Scenario 1 (external thyroid dose) | | | | | Exposure Scenario 2 (total thyroid dose) | | | | |
|---|---|---|---|---|---|---|---|---|---|---|
| Method | Conventional regression | | Bayesian method | | | Conventional regression | | Bayesian method | | |
| "True" slope (ERR/Gy) | Mean dose[a] | Median dose[b] | Original[c] | CM[d] | CMD[e] | Mean dose[a] | Median dose[b] | Original[c] | CM[d] | CMD[e] |
| 20 | 86.7 (91.3)[f] | 93.3 (96.6)[f] | 93.3 | 93.3 | 93.3 | 10.0 (12.1)[f] | 50.0 (52.2)[f] | 63.3 | 83.3 | 90.0 |
| 12 | 90.0 (95.3) | 93.3 (98.6) | 96.7 | 93.3 | 93.3 | 13.3 (12.6) | 50.0 (49.5) | 66.7 | 86.7 | 86.7 |
| 3 | 96.7 (99.9) | 96.7 (99.9) | 96.7 | 100 | 100 | 13.3 (11.5) | 53.3 (55.4) | 80.0 | 96.7 | 96.7 |
| Overall | 91.1 (95.5) | 94.4 (98.4) | 95.6 | 95.6 | 95.6 | 12.2 (12.1) | 51.1 (52.4) | 70.0 | 88.9 | 91.1 |

Conventional regression analysis using a single vector of 2,376 doses (one per subject)
  [a] individual mean doses
  [b] individual median doses

Bayesian Model Averaging using 4,999 cohort dose vectors of 2,376 doses (one per subject per vector)
  [c] simulated individual doses (Original)
  [d] conditional individual mean doses (CM)
  [e] conditional individual median doses (CMD)

[f] Parentheses denote inclusion percentage using all 5,000 simulated "true" disease status sets produced for each prescribed "true" slope of a linear dose-response for performance testing of conventional regression analysis methods.

Table 3. Relative Bias, Half-width of CI, and Relative Upper Bound of CI summarized from 90 performance tests using simulated "true" sets of disease status for Exposures Scenarios 1 and 2.

**Relative bias**

| | Scenario 1 (external thyroid dose) | | | | | Scenario 2 (total thyroid dose) | | | | |
|---|---|---|---|---|---|---|---|---|---|---|
| Method | Conventional regression | | Bayesian method | | | Conventional regression | | Bayesian method | | |
| "True" slope (ERR/Gy) | Mean dose[a] | Median dose[b] | Original[c] | CM[d] | CMD[e] | mean dose[a] | median dose[b] | Original[c] | CM[d] | CMD[e] |
| 20 | 0.15 | 0.14 | 0.14 | 0.13 | 0.13 | 0.62 | 0.35 | 0.38 | 0.30 | 0.27 |
| 12 | 0.14 | 0.14 | 0.15 | 0.13 | 0.14 | 0.66 | 0.39 | 0.38 | 0.29 | 0.28 |
| 3 | 0.18 | 0.20 | 0.17 | 0.18 | 0.20 | 0.70 | 0.65 | 0.42 | 0.35 | 0.32 |
| Overall | 0.16 | 0.16 | 0.16 | 0.15 | 0.16 | 0.66 | 0.46 | 0.40 | 0.31 | 0.29 |

**Half-width of CI**

| | Scenario 1 (external thyroid dose) | | | | | Scenario 2 (total thyroid dose) | | | | |
|---|---|---|---|---|---|---|---|---|---|---|
| Method | Conventional regression | | Bayesian method | | | Conventional regression | | Bayesian method | | |
| "True" slope (ERR/Gy) | Mean dose[a] | Median dose[b] | Original[c] | CM[d] | CMD[e] | Mean dose[a] | Median dose[b] | Original[c] | CM[d] | CMD[e] |
| 20 | 5.73 | 6.21 | 7.68 | 7.79 | 8.27 | 2.83 | 6.00 | 9.62 | 11.64 | 13.20 |
| 12 | 3.96 | 4.31 | 4.97 | 5.06 | 5.38 | 1.80 | 3.90 | 6.02 | 7.33 | 8.24 |
| 3 | 2.00 | 2.18 | 2.14 | 2.21 | 2.34 | 0.51 | 1.53 | 2.12 | 2.42 | 2.85 |
| Overall | 3.90 | 4.23 | 4.93 | 5.02 | 5.33 | 1.72 | 3.81 | 5.92 | 7.13 | 8.09 |

**Relative upper bound of CI (=UCL/Est.)**

| | Exposure Scenario 1 (external thyroid dose) | | | | | Exposure Scenario 2 (total thyroid dose) | | | | |
|---|---|---|---|---|---|---|---|---|---|---|
| Method | Conventional regression | | Bayesian method | | | Conventional regression | | Bayesian method | | |
| "True" slope (ERR/Gy) | Mean dose[a] | Median dose[b] | Original[c] | CM[d] | CMD[e] | Mean dose[a] | Median dose[b] | Original[c] | CM[d] | CMD[e] |
| 20 | 1.31 | 1.31 | 1.43 | 1.43 | 1.42 | 1.37 | 1.33 | 1.71 | 1.76 | 1.71 |
| 12 | 1.35 | 1.35 | 1.47 | 1.46 | 1.46 | 1.38 | 1.34 | 1.82 | 1.83 | 1.79 |
| 3 | 1.71 | 1.71 | 1.81 | 1.76 | 1.76 | 1.52 | 1.44 | 2.13 | 2.09 | 1.99 |
| Overall | 1.46 | 1.46 | 1.57 | 1.55 | 1.55 | 1.42 | 1.37 | 1.89 | 1.89 | 1.83 |

Conventional regression analysis using a single dose vector of 2,376 doses (one per subject)
  [a] individual mean doses
  [b] individual median doses

Bayesian Model Averaging using 4,999 cohort dose vectors of 2,376 doses (one per subject per vector)
  [c] simulated individual doses (Original)
  [d] conditional individual mean doses (CM)
  [e] conditional individual median doses (CMD)

# Figure 1.

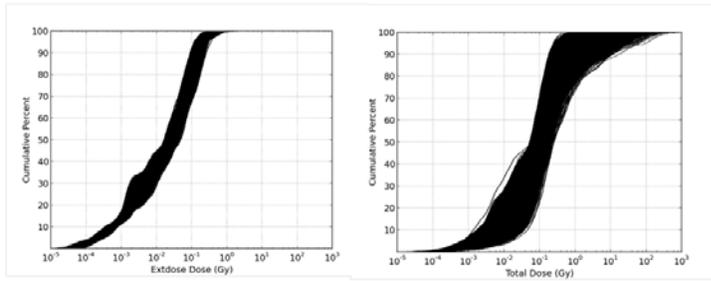

# Figure 2.

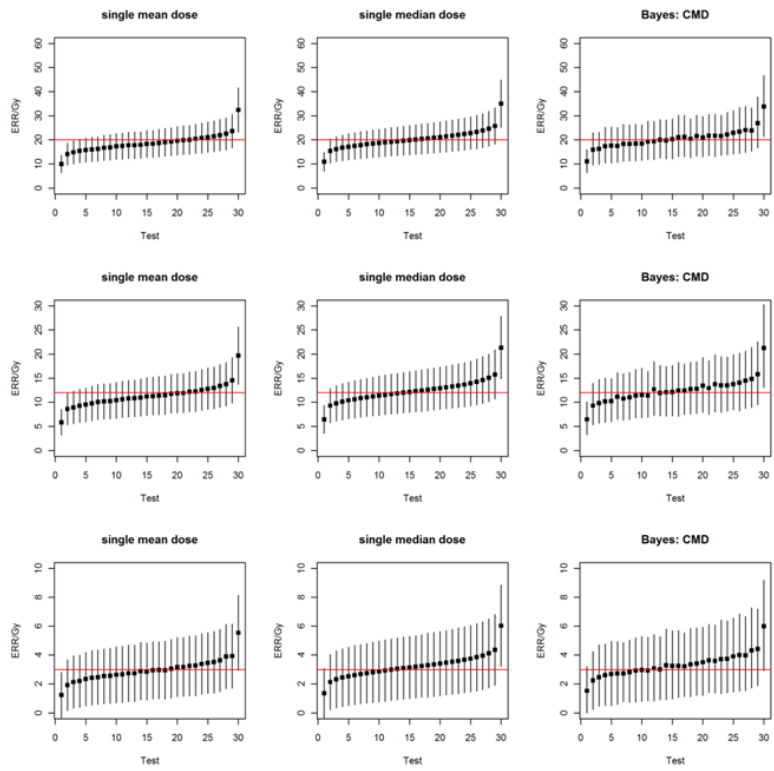

# Figure 3.

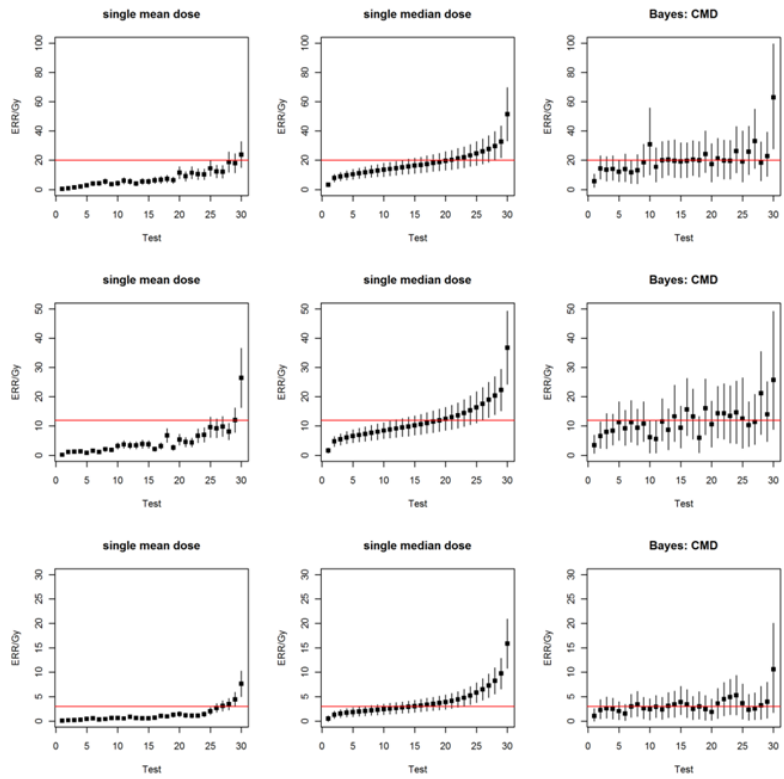

# Figure 4.

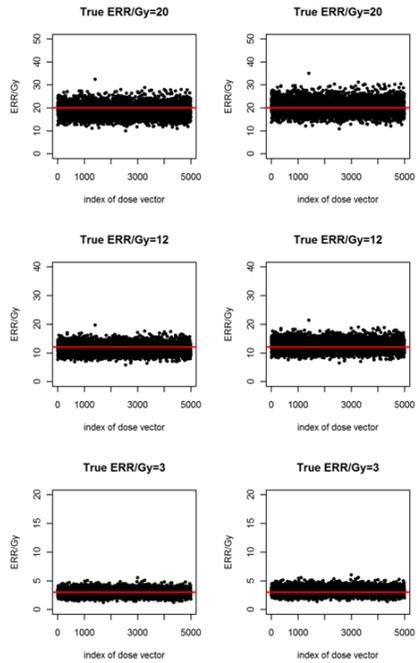

# Figure 5.

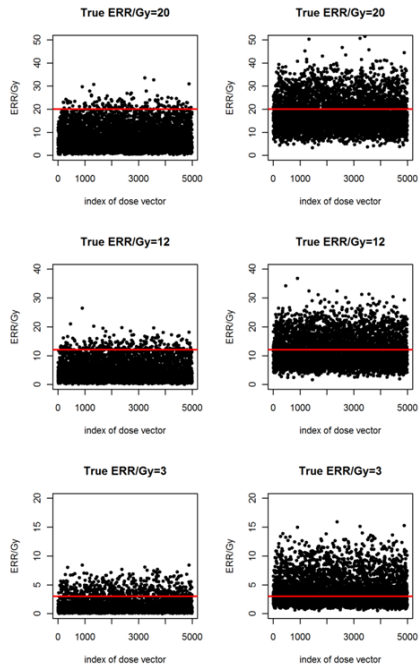

# Figure 6.

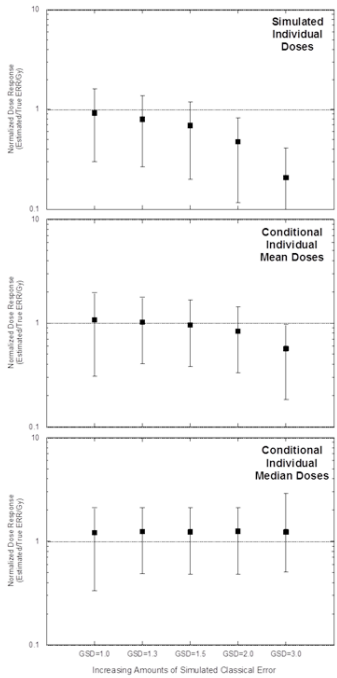

Figure S1

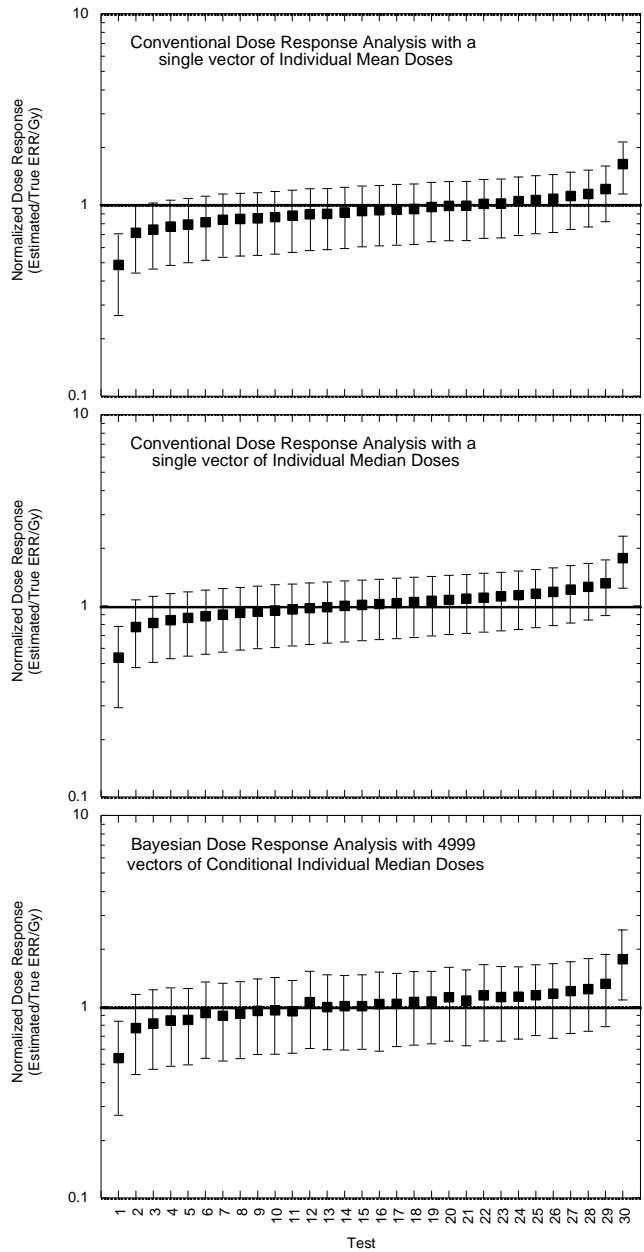

Figure S1. Results of 30 simulation tests for Scenario 1 (external thyroid dose) comparing the dose response analysis for conventional regression using a single vector of individual mean and individual median doses, with a Bayesian dose response using 4,999 vectors of conditional individual median doses. Results are normalized to a true ERR/Gy of 12 (i.e., results that reproduce the true ERR/Gy will equal 1.0).

Figure S2

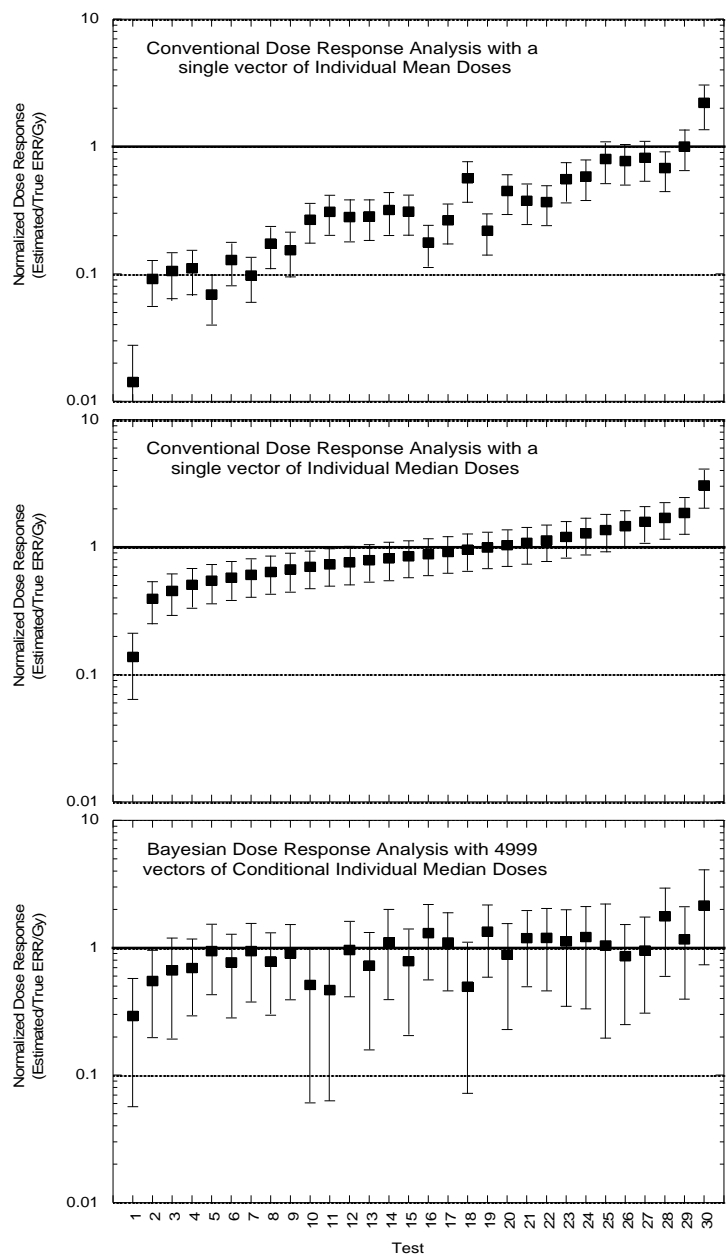

Figure S2. Results of 30 simulation tests for Scenario 2 (total thyroid dose) comparing the dose response analysis for conventional regression with a single vector of mean and median doses, with a Bayesian dose response using 4,999 vectors of conditional individual median doses. Results are normalized to a true ERR/Gy of 12 (i.e., results that reproduce the true ERR/Gy will equal 1.0).

**Appendix S1. Relationship between our problem and Bayesian model uncertainty paradigm**

Under our model uncertainty paradigm, we rely on Bernardo and Smith's (*34*) perspectives of *M*-closed, *M*-complete, and *M*-open for the relationship between the multiple realizations of dose vectors produced from the 2DMC algorithm and the true exposure model. The *M*-closed perspective is the situation in which the dose reconstruction method describes the 'true' exposure model exactly and thus is able to produce the true exposure in one of its multiple realizations of possibly true dose vectors. Under *M*-complete, the 'true' exposure model exists but the 2DMC algorithm is unable to reproduce the true dose vector exactly. However, a few of the 2DMC realizations of a possibly true dose vector should be reasonable approximations of the true exposure and, thus, we are able to evaluate and compare each of the alternative realizations of dose vectors. The *M*-open perspective is similar to the *M*-complete; the difference is that we cannot assume that the true exposure model is captured within the multiple realizations of dose vectors produced by the 2DMC dose estimation algorithm. In that case, we cannot develop any prior beliefs about the probability that each realization of 2DMC is equal to the true exposure model without having additional information. Most dose estimations for cohorts supporting epidemiologic studies are located somewhere between the extremes of the perspectives of *M*-complete and *M*-open, since we cannot guarantee that the dosimetry system generates the true dose vector.

**Appendix S2. Stochastic Approximation Monte Carlo (SAMC)**

We give a brief review of the SAMC method (*33*). We can reformulate the posterior distribution in terms of $U(\bullet)$, the energy function:

$$p(\underline{\alpha},\beta,\gamma|Data) = \exp(-U(\underline{\alpha},\beta,\gamma))/Z, (\underline{\alpha},\beta,\gamma) \in \Theta,$$

where $Z$ is the normalizing constant and $\Theta$ is the parameter space. Let $\theta$ denote a set of parameters, $\theta=(\alpha_1,\ldots, \alpha_J, \beta, \gamma)$ and $\theta \in \Theta$. When we write $U(\theta)=-\log(p_0(\theta|Data))$, where $p_0(\bullet)$ is the unnormalized posterior distribution, then $U$ is equivalent to the negative of the log-posterior distribution in Bayesian computation.

Without loss of generality we assume $\Theta$ is compact. In the implementation we set $\Theta=[-10^{100},10^{100}]$. We assume that the parameter space can be partitioned according to the energy function, U(θ), into S disjoint sub-regions denoted by $E_1=\{(\theta): U(\theta) \leq u_1\}$, $E_2=\{(\theta): u_1 < U(\theta) \leq u_2\}$,..., $E_{(S-1)}=\{(\theta): u_{(S-2)} < U(\theta) \leq u_{(S-1)}\}$, $E_S=\{(\theta): U(\theta) > u_{(S-1)}\}$, where $u_1,..,u_{(S-1)}$ are real numbers in increasing order and are specified by the user. The SAMC method tries to sample from each sub-region with a pre-specified frequency, $\underline{f}=(f_1,...,f_S)$ (e.g., equal frequency) using the trial distribution $p_w(\theta) \propto \sum_{s=1}^{S} \frac{f_s p_0(\theta)}{w_s} I(\theta \in E_s),$ where I(•) denotes an indicator function and $w_s = \int_{E_s} p_0(\theta) d\theta$ for s=1,...,S.

Since our problem involves parameter estimation with multiple models, (i.e., K linear dose-response models according to K dose vectors), the model space can be partitioned into K disjoint sub-regions ($E_1,..., E_K$). We use the SAMC model selection approach. We emphasize that our main interest is estimation of the slope of the linear dose response, β, (ERR $Gy^{-1}$), not the selection of different types of dose-response models with different shapes that depart from linear. We attempt to identify which among multiple dose vectors is a good approximate for the unknown true dose vector in this study. We illustrate how we implemented the SAMC method. Let K models denote $M_1, ..., M_K$, each of which are associated with one of the K dose vectors. The SAMC method consists of two stages: (1) Metropolis-Hastings (MH) sampling of θ and (2) weight updating. The weight, $\underline{\omega}=(\omega_1,...,\omega_K)$, denotes the working estimate of ($\log(w_1/f_1),...,\log(w_K/f_K)$) obtained at each iteration.

In MH sampling, we generate a sample $\theta^{(t)}$ from a Metropolis-Hastings kernel $K_{w^{(t)}}(\theta^{(t)},\bullet)$ with the proposal distribution $q(\theta^{(t)},\bullet)$ and the stationary distribution, $p_{\omega^{(t)}}(\theta) \propto \sum_{k=1}^{K} \frac{p_0(\theta)}{e^{\omega_k^{(t)}}} I(\theta \in E_k)$, where $\underline{\omega}^{(t)} = (\omega_1^{(t)},...,\omega_K^{(t)})$ at iteration t. Let $Q(M_i \to M_j)$ denote the proposed probability for a transition from model $M_i$ to model $M_j$. The proposed distribution satisfies irreducibility and aperiodicity for convergence.

First, generate dose vector selection parameter, $\gamma^*$, according to the proposal $Q(M_{\gamma^{(t-1)}} \to M_{\gamma^*})$ at iteration t. If $\gamma^*=\gamma^{(t-1)}$, then generate $\underline{\alpha}^*$ and $\beta^*$ from $p(\underline{\alpha},\beta | X, y, dose\ vector_{\gamma^{(t-1)}})$ by a single

MCMC iteration and set $(\gamma^t, \underline{\alpha}^t, \beta^t) = (\gamma^*, \underline{\alpha}^*, \beta^*)$. If $\gamma^* \neq \gamma^{(t-1)}$, then generate $\underline{\alpha}^*$ and $\beta^*$ from $p(\underline{\alpha}, \beta | X, y, dose\ vector_{\gamma^*})$ and accept $\gamma^*$, $\underline{\alpha}^*$ and $\beta^*$ with probability

$$\min\left\{1, \frac{e^{\omega_{M_{\gamma^{(t-1)}}}^{(t-1)}}}{e^{\omega_{M_{\gamma^*}}^{(t-1)}}} \frac{p(\alpha^*, \beta^* | X, y, dose\ vector_{\gamma^*})}{p(\alpha^{(t-1)}, \beta^{(t-1)} | X, y, dose\ vector_{\gamma^{(t-1)}})} \frac{Q(M_{\gamma^*} \to M_{\gamma^{(t-1)}})}{Q(M_{\gamma^{(t-1)}} \to M_{\gamma^*})}\right\}.$$

In the weight updating stage, $\underline{\omega}^* = \underline{\omega}^{(t-1)} + \delta^t(\underline{e}^t - \underline{f})$, where $\delta^t$ denotes the gain factor sequence and $\underline{e}^t = (e_1^{(t)}, \ldots, e_K^{(t)})$ and $e_k^{(t)} = 1$ if $\gamma^{(t)} = k$ and 0 otherwise. The gain factor sequence should be a positive, non-decreasing sequence satisfying the following conditions: $\sum_{t=0}^{\infty} \delta^t = \infty$ and $\sum_{t=0}^{\infty} (\delta^t)^\nu < \infty$, for some $\tau \in (1,2)$. More details on the implementation of SMAC are described in Liang et al. (*33*).